\documentclass[twocolappendix]{emulateapj}

\usepackage{soul,color,amsmath}

\shorttitle{The Boosted Fireball Model}
\shortauthors{Duffell \& MacFadyen}

\begin{document}

\title{A ``Boosted Fireball" Model for Structured Relativistic Jets}

\author{Paul C. Duffell and Andrew I. MacFadyen}
\affil{Center for Cosmology and Particle Physics, New York University}
\email{pcd233@nyu.edu, macfadyen@nyu.edu}

\begin{abstract}

We present a model for relativistic jets which generates a particular angular distribution of Lorentz factor and energy per solid angle.  We consider a fireball with specific internal energy $E/M$ launched with bulk Lorentz factor $\gamma_B$.  In its center-of-momentum frame the fireball expands isotropically, converting its internal energy into radially expanding flow with asymptotic Lorentz factor $\eta_0 \sim E/M$.  In the lab frame the flow is beamed, expanding with Lorentz factor $\Gamma = 2 \eta_0 \gamma_B$ in the direction of its initial bulk motion and with characteristic opening angle $\theta_0 \sim 1/\gamma_B$.  The flow is jet-like with $\Gamma \theta_0 \sim 2 \eta_0$ such that jets with $\Gamma > 1/\theta_0$ are naturally produced.  The choice $\eta_0 \sim \gamma_B \sim 10$ yields a jet with $\Gamma \sim 200$ on-axis and angular structure characterized by opening angle $\theta_0 \sim 0.1$ of relevance for cosmological GRBs, while $\gamma_B \gtrsim 1$ may be relevant for low-luminosity GRBs.
The model produces a family of outflows, of relevance for different relativistic phenomena with structures completely determined by $\eta_0$ and $\gamma_B$. We calculate the energy per unit solid angle for the model and use it to compute light curves for comparison with the widely used top-hat model.  The jet break in the boosted fireball light curve is greatly subdued when compared to the top-hat model because the edge of the jet is smoother than for a top-hat.  This may explain missing jet breaks in afterglow light curves.

\end{abstract}


\section{Introduction}
\label{sec:intro}

Relativistic, collimated outflows are observed in many contexts in astrophysics, most notably gamma ray bursts, whose outflows can have Lorentz factors in the hundreds, and  may be collimated to within opening angles of a few degrees \citep{piran04,zm04}.  In order to determine the properties of these jets, one must first choose a model for the jet structure since little can be said definitively from first principles about the distribution of energy and momentum in the outflow.

Numerical calculations of jet dynamics must assume a model for the initial conditions.  One can only trust the results of such calculations to the extent that one trusts the initial conditions.  This problem is compounded by causality.  Because fluid elements on relativistic trajectories cannot communicate with anything outside of an opening angle $\Delta \theta \sim 1/\Gamma$, where $\Gamma$ is a characteristic Lorentz factor of the outflow, angular fluctuations in the jet do not wash out until the jet has decelerated to a sufficiently low Lorentz factor.

Nevertheless, \emph{some} model for the outflow is needed, and a simple choice is to assume a spherically symmetric solution, then truncate the profile at some opening angle.  This choice has been employed by most authors \citep{r99,s99,pm99,mod00,kp00,gr01,gr02,zm09,hendrik10,wygoda11,hendrik11,decolle12}, but a jet with some other angular structure would have different observational properties \citep{ros02,ros04}.  Moreover, certain seemingly minor changes to the jet structure, like choosing for the outflow velocity to be parallel to the axis instead of radial, can result in completely different jet dynamics \citep{gruz07}.

More detailed jet models have been proposed and utilized (for a review see \cite{granrev}), for example assuming a Gaussian or power-law profile for the energy per solid angle as a function of angle from the jet axis \citep{zm02,kg03}.  Such models are not physically motivated, but are parameterized ``guesses" which can be fit to data.  More exotic but physically motivated models for jet structure have also been proposed \citep{le93,le00,th05,lb02,peng05}, however these models are generally designed to describe some particular feature of a given engine or to model observed light curves features \citep{berger03}.  So far, there is a lack of a simple, physically-motivated model for generic relativistic outflows whose opening angle is not imposed in an ad-hoc fashion.

Here we propose a jet model whose opening angle arises naturally, and is not presented as a truncation.  Rather, it generates a parameterized family of models, with a spherical fireball at one extreme, and a completely directed parallel flow at the other.  From this model we predict a particular angular structure for relativistic jets.  The model we propose is a ``boosted fireball".

\begin{figure}
\epsscale{1.0}
\plotone{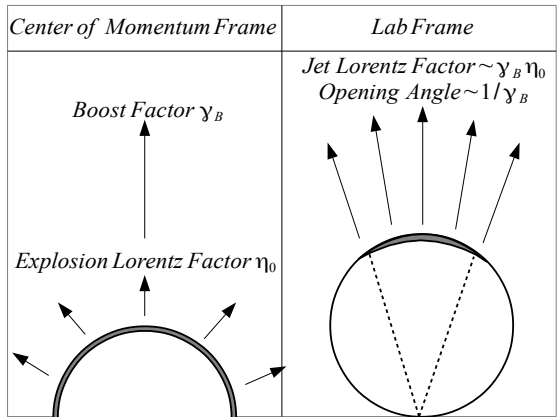}
\caption{ Diagram showing the basic features of the boosted fireball model.  The fireball is spherical in its center-of-momentum frame, and beamed with a characteristic opening angle $\sim 1/\gamma_B$ in the lab frame.
\label{fig:cartoon} }
\end{figure}

The basic idea is a simple modification of the fireball model \citep{p86,g86} in which the fireball is launched with a bulk Lorentz factor $\gamma_B$ from the central engine.  First we deposit an internal energy E into a mass M, in the fireball's rest frame.  We then view the flow in a reference frame which is moving with respect to this center-of-momentum frame, with boost factor $\gamma_B$.  If the outflow attains a Lorentz factor of $\eta_0$ in its center-of-momentum frame, in the ``boosted" frame (hereafter ``lab frame"), it will have characteristic Lorentz factor $\Gamma \sim \gamma_B \eta_0$ and opening angle $\theta_0 \sim 1/\gamma_B$ (Figure \ref{fig:cartoon}).  In the limit $\gamma_B \rightarrow 1$, the solution is a standard spherically symmetric fireball.  In the limit $\eta_0 \rightarrow 1$, the solution is a directed flow with negligible opening angle.

This model provides a language with which to rule out certain types of flows.  It can be employed as initial conditions for afterglow light curve calculations, which can be used to study the parameter space of $\gamma_B$ and $\eta_0$ to determine, for example, which regions of the parameter space are consistent with afterglow data.

\section{Phenomenology}
\label{sec:phen}

The motivation for this model comes from the physics of collapsar jets, which have been studied numerically \citep{mw99,aloy00,mwh01,zw03,zw04,mlb07,sasha08,lopez13,mi13}, and modeled analytically \citep{bnps11}.  First, a tunnel is drilled through the progenitor, aided by a hot cocoon which recollimates the relativistic stream of matter.  This tunnel acts like the barrel of a gun, through which hot, relativistic matter is ejected after experiencing internal and recollimation shocks and subsequent expansions. Energy is repeatedly converted back and forth between internal and kinetic forms during this process.  During the operation of the central engine a train of blobs can be ejected from the star as the relativistic flow repeatedly interacts with the hot turbulent cocoon which modulates it.
In many cases, matter is ejected with Lorentz factors of a few tens, but post-shock acceleration increases the Lorentz factor into the hundreds.  The process of post-shock acceleration can be viewed in the center-of-momentum frame of the ejecta, in which the hot matter is envisioned as a fireball, expanding isotropically outward and converting the internal energy of the ejecta into the kinetic energy of radial outflow.

In the lab frame, this spherical outflow is beamed in the direction of the boost, with characteristic opening angle which scales as the inverse of the boost Lorentz factor.  This can be understood from very simple considerations; in its own frame, the width of the blast is $R = \tau$ (assuming ultrarelativistic outflow for the moment), where $\tau$ is the elapsed time in this frame, and using units for which $c=1$ henceforth.  In the lab frame, the transverse width is the same but the fireball has propagated a distance $d = t = \gamma_B \tau$.  Thus, the characteristic opening angle is $\theta_0 \sim R/d = 1/\gamma_B$.  This line of reasoning does not account for simultaneity effects, which will modify some of these expressions, but the intuition is generally correct in that most of the energy is collimated into opening angle $1/\gamma_B$.

\section{Calculation of Jet Structure}
Before describing the boosted fireball, we first review the basic properties of a standard, non-boosted fireball.  This case has been previously investigated both analytically and numerically \citep{extra1,extra2,kps99}.

\begin{figure*}
\epsscale{1.0}
\plotone{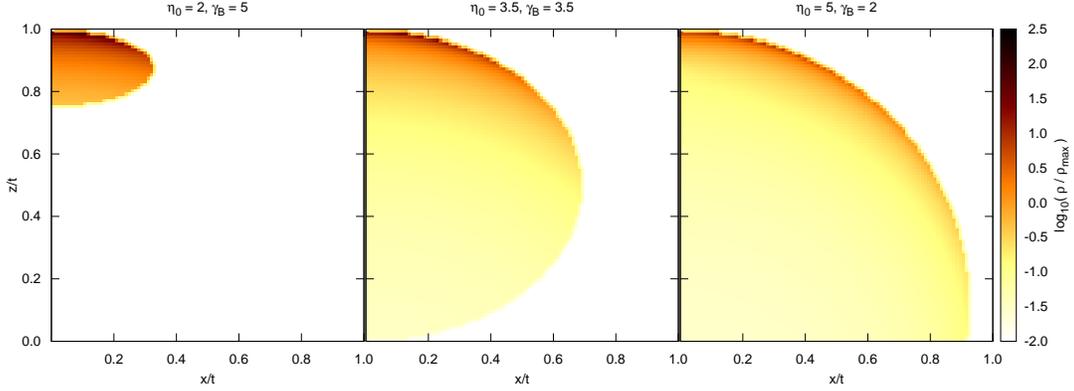}
\caption{ Several examples of boosted fireball solutions, for different values of $\eta_0$ and $\gamma_B$.  We plot logarithm of density, as given by equations (\ref{eqn:appxrho}) and (\ref{eqn:boostrho}).  Moderate Lorentz factors were chosen to make visualization clearer.
\label{fig:2d} }
\end{figure*}

\subsection{The Fireball Model}
\label{sec:fire}

Deposit an energy E and mass M within some radius $\Delta_0$.  
First, the fireball explodes due to its overpressure, undergoing an acceleration phase during which its internal energy is converted to kinetic energy.  Eventually, the flow becomes cold and attains its maximum Lorentz factor $\eta_0 \sim E/M$.  This occurs at time $t \sim \Delta_0 \eta_0$.  The width $\Delta$ of the outgoing shell is still equal to $\Delta_0$ at this time, but eventually at time $t \sim \Delta_0 \eta_0^2$, variations of velocity in the shell cause it to spread, keeping the ratio $\Delta/t$ fixed at $\sim 1/\eta_0^2$.  After this time, the outflow becomes self-similar, and the velocity field is a Hubble flow. 
For now we assume the surrounding medium is of neglibly small density so that the amount of swept up mass is negligible, and does not decelerate the shell.  (Eventually this will change after the shell has expanded to a large enough radius.)

The resulting outflow satisfies:

\begin{equation}
\rho_{sph}( r , t ) = \rho_0( r/t ) ( t_0 / t )^3
\end{equation}
\begin{equation}
\vec v( r , t ) = \left\{ \begin{array}
				{l@{\quad \quad}l}
				\vec r / t & r < R  		\\  
    			0 & \text{otherwise}  		\\
    			\end{array} \right.
    			\label{eqn:hubble}
\end{equation}
\begin{equation}
P \ll \rho
\end{equation}

where $R = v_{0} t$ is the radius of the outflow (with $v_{0} = \sqrt{1 - 1/\eta_0^2}$ the maximum velocity attained in the flow), and $t_0$ is some fiducial time.

In principle, the density profile in unspecified, as it depends on exactly how the mass was distributed in the initial explosion, but in practice generically $\rho$ is described by a thin shell with width $\Delta \sim t/\eta_0^2$, just as the local fluid Lorentz factor is, by equation (\ref{eqn:hubble}).  This profile could be modeled very simply with a top-hat:

\begin{equation}
\rho_0(r/t) = \left\{ \begin{array}
				{l@{\quad \quad}l}
				\rho_1 & v_{0} - 1/\eta_0^2 < r/t < v_{0}  		\\  
    			0 & \text{otherwise}  							\\
    			\end{array} \right. 
\end{equation}
where $\rho_1$ is chosen so that the total mass $\int 4 \pi r^2 \gamma(r) \rho(r) dr = M$.

In practice, we find in numerical calculations \citep{dm13} that the density profile can be reasonably well described by the following profile:
\begin{equation}
\rho(r,t) = \rho_{max} \left( {1 - R/t \over 1 - r/t} \right),
\label{eqn:appxrho}
\end{equation}
where 
\begin{equation}
\rho_{max} = {\eta_0 M \over 4 \pi t^3}.
\label{eqn:rhomax}
\end{equation}

Eventually, when this outflow sweeps up enough mass ($\sim M/\eta_0$), the shell begins to decelerate, and its energy is transferred to a blastwave which is described by the Blandford-McKee solution \citep{bm76}.

\subsection{The Boosted Fireball} 
\label{sec:boost}

A very simple extension of this idea is to take the solution above as a function of space and time, and perform a Lorentz boost by some factor, $\gamma_B$.  If the unboosted explosion has energy $\sim \eta_0 M$, the boosted explosion has energy $\sim \eta_0 \gamma_B M$.  The solution (before interacting with the circumburst medium) is

\begin{equation}
\rho( x , y , z , t ) = \rho_{sph}( x , y , \tilde z , \tilde t )
\label{eqn:boostrho}
\end{equation}
\begin{equation}
\vec v( r , t ) = \left\{ \begin{array}
				{l@{\quad \quad}l}
				\vec r / t & \text{(interior)} 		\\  
    			0 & \text{(exterior)}  				\\
    			\end{array} \right.
\end{equation}
where $\tilde z$ and $\tilde t$ are the boosted variables:
\begin{equation}
\tilde z = \gamma_B ( z - v_B t)
\end{equation}
\begin{equation}
\tilde t = \gamma_B ( t - v_B z).
\end{equation}
The equation for velocity follows from self-similarity, and from the Lorentz invariance of Hubble flows.  We show a few examples of such flows in Figure \ref{fig:2d}.

The structure of this flow can be found analytically by calculating the maximum Lorentz factor as a function of $\theta$.  This can be done most easily using four-vectors.  Define $w^{\mu}$ to be the four-velocity associated with the boost (i.e. the velocity of the center of mass), while $u^{\mu}_{max}$ is the local fluid four-velocity at the front of the outflow.  Then, we can equate their  scalar product in the two different frames:
\begin{equation}
- w_{\mu} u_{max}^{\mu} = \eta_0 = \gamma_B \gamma_{max} ( 1 - v_B v_{max} cos \theta )
\end{equation}
From this, the maximum Lorentz factor as a function of angle can be reconstructed:
\begin{equation}
\gamma_{max}(\theta) = \gamma_B {\eta_0 + v_B cos \theta \sqrt{ \eta_0^2 v_0^2 - \gamma_B^2 v_B^2 sin^2 \theta } \over  1 + \gamma_B^2 v_B^2 sin^2 \theta  }
\label{eqn:gammax}
\end{equation}
This formula can immediately be used to describe several features of the outflow.  First, if $\eta_0 < \gamma_B$, then there exists an angle $\theta_{max}$ for which the argument of the square root changes sign.  This signifies that $100\%$ of the ejecta is contained within $\theta_{max}$, so that the flow resembles a relativistic ``blob" of matter (as in the left panel of Figure \ref{fig:2d}).  $\theta_{max}$ can be simply calculated by setting the square root to zero:
\begin{equation}
sin( \theta_{max} ) = { \eta_0 v_0 \over \gamma_B v_B }
\end{equation}
However, $\theta_{max}$ is not necessarily the opening angle of the jet.  This can be defined by $\theta_{1/2}$, the angle at which $\gamma_{max}(\theta)$ attains its half-maximum.  For example, if $\eta_0$ and $\gamma_B$ are ultra-relativistic, then this is attained when the denominator of Equation (\ref{eqn:gammax}) equals 2:
\begin{equation}
sin( \theta_{1/2} ) \rightarrow { 1 \over \gamma_B }.
\end{equation}
The opening angle of the jet is the minimum of these two angles:
\begin{equation}
\theta_0 = min( \theta_{1/2} , \theta_{max} )
\end{equation}
The case of an unboosted fireball is attained in the limit $\gamma_B \rightarrow 1$.  In this case,
\begin{equation}
\gamma_{max}(\theta) \rightarrow \eta_0 
\end{equation}
A directed outflow is found by choosing $\eta_0$ close to unity, i.e. a nonrelativistic fireball in the boosted frame.  In this limit:
\begin{equation}
\gamma_{max}(\theta) \rightarrow \left\{ \begin{array}
				{l@{\quad \quad}l}
				\gamma_B & \theta < v_0/\gamma_B 	\\  
    			1 & \text{otherwise}				\\
    			\end{array} \right.
\end{equation}
Another simplification can be found taking the intermediate case $\gamma_B = \eta_0$.  In this case, the structure of the flow simplifies to
\begin{equation}
\gamma_{max}(\theta) = { 1 + v_B^2 cos^2 \theta  \over 1 - v_B^2 cos^2 \theta }
\label{eqn:same}
\end{equation}
If we wish to find how energy is distributed as a function of opening angle, we must first find the maximum density.  This, however, is a simple matter as the angular structure of density is purely determined by relative simultaneity; we need only consider how $\rho_{max}$ scales with time:
\begin{equation}
\rho_{max}(\theta,t) = \rho_{max}(\tilde t) \propto {\tilde t}^{-3}.
\end{equation}
At the front of the flow, the transformation of time simply modifies it by the following factor:
\begin{equation}
\tilde t = t ( \eta_0 / \gamma_{max} ).
\end{equation}
Since the angular structure for density is purely given by the angular structure for $t$, we can immediately write down the formula:
\begin{equation}
\rho_{max}(\theta,t) = \rho_{max}(t) \left( {\gamma_{max}(\theta) \over \eta_0 } \right)^3.
\label{eqn:density}
\end{equation}

Now that we know how the density and Lorentz factor vary with $\theta$, we can compute how the energy per solid angle varies with $\theta$.  For a cold flow with negligible thermal energy,

\begin{equation}
{dE \over d\Omega} = \int \gamma^2 \rho r^2 dr 
\end{equation}

\begin{equation}
\sim \gamma_{max}^2 \rho_{max} r^2 \Delta 
\end{equation}

\begin{equation}
\sim \rho_{max}(\theta,t) t^3
\label{eqn:escale}
\end{equation}

Using the scaling (\ref{eqn:escale}) along with Equation (\ref{eqn:density}) taking $\rho_{max}(t) \propto t^{-3}$, we can derive a complete solution for a given isotropic equivalent energy $E_{iso}$:

\begin{figure}
\epsscale{1.0}
\plotone{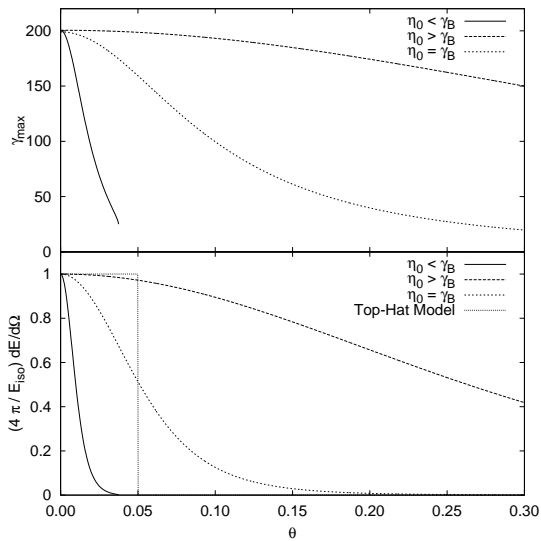}
\caption{ Lorentz factor (Eq. \ref{eqn:gammax}) and energy per solid angle (Eq. \ref{eqn:eprofile}) are plotted as a function of $\theta$, for a few different models.  All models here are chosen to have a peak Lorentz factor of 200.  Solid curves have $\eta_0 = 2.13$ and $\gamma_B = 50$.  Dashed curves have $\eta_0 = 50$ and $\gamma_B = 2.13$.  Dotted curves use $\eta_0 = \gamma_B = 10$.
\label{fig:js} }
\end{figure}

\begin{equation}
{dE \over d\Omega} = {E_{iso} \over 4 \pi} \left({\gamma_{max}(\theta) \over \gamma_{max}(0) }\right)^3 
\label{eqn:eprofile}
\end{equation}

\section{Observational Effects}
\label{sec:obs}	

A proper study of the observational predictions of the boosted fireball model would entail detailed hydrodynamical calculations in which the outflow interacts with the circumburst medium.  We will attempt this in a future study, but for now we can perform a basic calculation demonstrating some observational differences between our model and what has been assumed previously.

After this outflow has swept up enough mass, it will heat up and begin to decelerate.  During this time the blastwave produces synchrotron radiation, generating the observable afterglow.  To determine the correct structure of this blastwave would require a proper numerical calculation, but it is reasonable to assume that the blastwave will inherit $dE/d\Omega$ from the outflow which produced it.  The reason this is a reasonable assumption is that causality prevents energy from being redistributed on angular scales larger than $\Delta \theta \sim 1/\gamma_{max}$, so at worst there will probably be a small amount of smoothing on those angular scales.  If $\eta_0 \gg 1$, the opening angle of the jet is much larger than $1/\gamma_{max}$, so such a redistribution of energy is probably negligible.

\begin{figure}
\epsscale{1.0}
\plotone{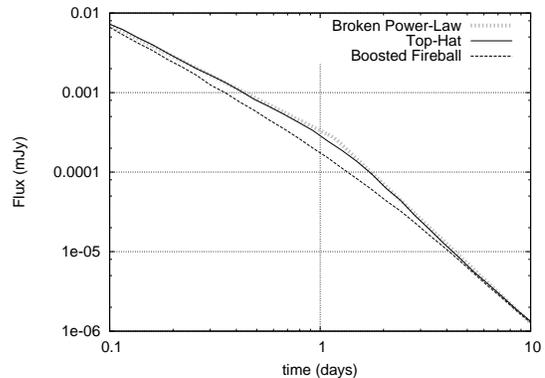}
\caption{ Afterglow light curves from two different models, a top-hat Blandford-McKee model and the boosted fireball model.  Both models assume isotropic equivalent energy $E_{iso} = 4 \times 10^{53}$ ergs and circumburst number density $n_{ISM} = 1/cm^3$.  The top-hat model is truncated at opening angle $\theta_0 = 0.1$.  The boost for the fireball was chosen to match the late-time afterglow, $\gamma_B = 6.3 = 0.63/\theta_0$.  $dE/d\Omega$ is calculated from (\ref{eqn:eprofile}), assuming $\gamma_0 = \eta_0$, meaning the angular profile for Lorentz factor simplifies to (\ref{eqn:same}).  Both light curves have asymptotic behavior which matches a broken power-law, but the top-hat model exhibits a sharper jet break.
\label{fig:lc} }
\end{figure}

This provides an angular structure, and for the radial structure we assume a Blandford-McKee profile \citep{bm76}.  This too is reasonably well-motivated by causality arguments, but more importantly the emission should for the most part depend on the shock jump conditions and the width of the energy-containing region behind the shock, both of which should agree with Blandford-McKee.  The isotropic equivalent energy generating the Blandford-McKee profile is taken as a function of angle $E_{iso}(\theta)$, chosen such that $dE/d\Omega$ matches the profile of Equation \ref{eqn:eprofile}.  Assuming that this flow persists and does not spread, we calculate the synchrotron radiation produced, using the same methods as in \cite{hendrik09,hendrik10,hendrik11}, with a very simplified radiation model (no synchrotron self-absorption).  For comparison, we perform the same calculation using a ``top-hat" profile for $dE/d\Omega$.  Results are shown in Figure \ref{fig:lc}.

The top-hat model shows a clear break in the light curve signifying that the observer has begun to see the edge of the jet \citep{fruc99,kul99}.  The boosted fireball model also has a jet break, but it is smoothed a great deal, owing to the fact that the jet does not have sharp edges in this model.  Such a light curve might be smooth enough for the jet break to go unnoticed in the afterglow data, a possible explanation for some missing jet breaks \citep{bur06,rac09}.  

\section{Discussion}
\label{sec:disc}

We expect the boosted fireball model to be more realistic than a top-hat model, simply because the jet has no reason to have sharp edges.  This by itself makes a noticeable difference in afterglow light curve predictions.

The model could also be modified by relaxing the assumption that the outflow has attained a self-similar state by the time it begins to decelerate.  In this case, another parameter should be added to the model, for example the thickness of the shell at the deceleration time.

The boosted fireball model can more generally be used to describe a variety of relativistic outflows.  Perhaps low-luminosity gamma ray bursts have significant $\eta_0$ but small $\gamma_B$, which would imply that the explosion was not ejected as rapidly from the progenitor, perhaps because jetted flow did not escape the stellar surface before the central engine died \citep{macfadyen2000,mw01}.  Shock breakout might take an outflow which was otherwise destined to be spherical and provide collimation with a boost.  On the opposite end of parameter space, a much more directed outflow requires $\eta_0 \ll \gamma_B$, which would in this context require an engine capable of accelerating the ejecta without heating it.  The boosted fireball might also be applicable to jets produced in active galactic nuclei or in compact binary mergers.

The case $\gamma_B = \eta_0 = 10$ produces a jet with Lorentz factor $\Gamma = 200$ and opening angle $\theta_0 = 0.1$, which is appropriate for cosmological gamma ray bursts.  In the context of this model, this suggests some mechanism in the engine for bringing kinetic and thermal energy into approximate equipartition, e.g. a hot cocoon.

\acknowledgments
This research was supported in part by NASA through grant NNX10AF62G issued through the Astrophysics Theory Program and Chandra grant TM3-14005X and by the NSF through grant AST-1009863.  

We are grateful to Andrei Gruzinov and Hendrik van Eerten for helpful comments and discussions.  We thank the referee Peter M{\'e}sz{\'a}ros for his helpful review.

{}

\end{document}